\documentstyle[preprint,aps,epsfig]{revtex}
\tightenlines
\begin{document}
\draft
\preprint{\vbox{\hbox{IFT--P.081/2000}}}
\title{Searching for Anomalous Higgs Couplings in Peripheral Heavy Ion
Collisions at the LHC. }
\author{ S.\ M.\ Lietti, A.\ A.\ Natale, C.\ G.\ Rold\~ao and R.\ Rosenfeld}
\address{Instituto de F\'{\i}sica Te\'orica,
Universidade  Estadual Paulista, \\
Rua Pamplona 145, CEP 01405-900 S\~ao Paulo, Brazil.}
\date{\today}
\maketitle
\widetext
\begin{abstract}
We investigate the sensitivity of the heavy ion mode of the LHC to
anomalous Higgs boson couplings to photons, $H\gamma\gamma$, through
the analysis of the processes  $\gamma \gamma \to b \bar{b}$ and 
$\gamma \gamma \to \gamma\gamma$ in peripheral heavy ion collisions. 
We suggest cuts to improve the signal over background ratio and determine
the capability of LHC to impose bounds on anomalous couplings
by searching for a Higgs boson signal in these modes. 
\end{abstract}


\section{Introduction}
\label{sec:int}

The Higgs boson is the only particle in the Standard Model (SM) that
has not yet been confirmed experimentally. It is responsible for the
mass generation of fermions and gauge bosons. The search for the Higgs
boson is the main priority in high energy experiments and hints of its
existence may have been already seen at LEP\cite{higgs} at around
$m_H \sim 115$ GeV. However, once found the detailed study of its
couplings could give information on the mass generation mechanism and
on physics beyond the Standard Model.

An intermediate-mass Higgs boson could also be produced in peripheral
heavy ion collisions  through photon-photon interactions
\cite{baur,baur2}. This possibility,  in the context of the SM, has 
been explored in detail in the literature \cite{papa,cahn,muller},
with the general conclusion that the chances of finding the standard
model Higgs in the photon-photon case are marginal.

However, the Standard Model is only an effective low energy theory of
a more complete model and one expects deviations from its predictions.
A convenient way  to parameterize deviations of the Standard Model
predictions is the effective theory approach \cite{effective}. In this
scenario, we assume that the existence of new physics, associated
with a high--energy scale $\Lambda$, can manifest itself at low
energies via the process of integrating-out heavy degrees of freedom.
These effects are then described by effective operators involving the
spectrum of particles belonging to the low--energy theory. At this
point we have two possibilities: either the Higgs boson is light and
it should be included in the effective operators or the Higgs boson is
heavy and should also be integrated out. In this work we will adopt
the former possibility, where the gauge group $SU(2)_L \otimes U(1)_Y$ is
linearly realized. In this case, the effective lagrangian will
generate anomalous Higgs couplings.

In this Letter we explore the capabilities of peripheral heavy ion
collisions in constraining anomalous Higgs couplings, which could in
principle arise from new physics beyond the SM. We analyse the processes
$\gamma \gamma \to b \bar{b},\gamma\gamma$. After simulating the signal and
background, we find optimal cuts to maximize their ratio. We show how
to use energy and invariant mass spectra of the final state $b \bar{b}$
or photon pair in order to identify the presence of a Higgs boson and
extract information about its couplings. Finally, we compare the
bounds on the anomalous couplings that will be possible to extract
from our analyses to bounds coming from other processes in different
machines.

\section{Anomalous Higgs Couplings and Effective Lagrangians}
\label{sec:eff}

In the linear representation of the $SU(2)_L \otimes U(1)_Y$ symmetry
breaking mechanism, the SM model is the lowest order approximation
while the first corrections, which are of dimension six, can be
written as
\begin{equation}
{\cal L}_{\text{eff}} = \sum_n \frac{f_n}{\Lambda^2} {\cal O}_n \;\; ,
\label{l:eff}
\end{equation}
where the operators ${\cal O}_n$ involve vector--boson and/or
Higgs--boson fields with couplings $f_n$ \cite{linear}. This
effective Lagrangian describes the phenomenology of models that
are somehow close to the SM since a light Higgs scalar doublet is
still present at low energies. Of the eleven possible operators
${\cal O}_{n}$ that are $P$ and $C$ even, only three of them modify
the Higgs--boson couplings to photons \cite{dim6:zep,Hagiwara2},
\begin{eqnarray}
&&{\cal O}_{BW} =  \Phi^{\dagger} \hat{B}_{\mu \nu}
\hat{W}^{\mu \nu} \Phi \;\; , \nonumber \\
&&{\cal O}_{WW} = \Phi^{\dagger} \hat{W}_{\mu \nu}
\hat{W}^{\mu \nu} \Phi  \;\; , \label{eff} \\
&&{\cal O}_{BB} = \Phi^{\dagger} \hat{B}_{\mu \nu}
\hat{B}^{\mu \nu} \Phi
 \;\; , \nonumber
\end{eqnarray}
where $\Phi$ is the Higgs doublet,
$\hat{B}_{\mu \nu} = i (g'/2) B_{\mu \nu}$, and $\hat{W}_{\mu \nu} =
i (g/2) \sigma^a W^a_{\mu \nu}$, with $B_{\mu \nu}$ and $ W^a_{\mu
\nu}$ being respectively the $U(1)_Y$ and $SU(2)_L$ field strength
tensors. In the unitary gauge, the anomalous $H\gamma\gamma$ coupling
is given by
\begin{equation}
{\cal L}_{\text{eff}}^{\text{HVV}} =
g_{H \gamma \gamma} \; H A_{\mu \nu} A^{\mu \nu} \;\; ,
\end{equation}
where $A_{\mu \nu} = \partial_\mu A_\nu - \partial_\nu
A_\mu$ and
\begin{equation}
g_{H \gamma \gamma} = - \left( \frac{g M_W}{\Lambda^2} \right)
                       \frac{s^2 (f_{BB} + f_{WW} - f_{BW})}{2} \;\; ,
\label{coupling}
\end{equation}
with $g$ being the electroweak coupling constant and $s \equiv
\sin \theta_W$.

The operator ${\cal O}_{BW}$ contributes at tree level to the
vector--boson two--point functions, and consequently is severely
constrained by low--energy data \cite{rujula,dim6:zep}. The present
95\% CL limits on these operators for 90 GeV $ \le m_H \le$ 800 GeV
and $m_{\text{top}}=175$ GeV read
\cite{hms},
\begin{equation}
-1.0 \le \frac{f_{BW}}{\Lambda^2} \le 8.6 \hbox{ TeV}^{-2} \;\; .
\label{b:bw}
\end{equation}

The remaining operators can be indirectly constrained via their
one--loop contributions to low--energy observables, which are
suppressed by factors $1/(16 \pi^2)$. Using the ``naturalness''
assumption that large cancellations do not occur among their
contributions, we can consider only the effect of one operator at
a time. In this case, the following constraints at 95\% CL
(in units of TeV$^{-2}$) arise \cite{hms}
\begin{equation}
 -24\leq \frac{f_{WW}}{\Lambda^2}\leq 14\;\;\;\; , \;\;\;\;
 -79\leq \frac{f_{BB}}{\Lambda^2}\leq 47\; .  \label{blindlim}
\end{equation}
These limits depend in a complex way on the Higgs mass. The values
quoted above for the sake of illustration were obtained for
$M_{H}=200$ GeV.

There are also limits coming from direct Higgs searches at LEP II
\cite{ours:lep2aaa}, Tevatron \cite{ours:tevatron} colliders.  The
combined analysis \cite{ours:comb} of these signatures yields the
following 95\% CL bounds on the anomalous Higgs interactions (in
TeV$^{-2}$):
\[
-7.5 \leq  \frac{f_{WW(BB)}}{\Lambda^2} \leq 18
\]
for $m_H\leq 150$ GeV. These limits can be improved by a factor 2--3
in the upgraded Tevatron runs. The 95\% CL bounds on the anomalous
Higgs interactions (in TeV$^{-2}$) coming from direct Higgs
searches via gluon gluon fusion at LHC \cite{ours:lhc} collider are
\begin{equation}
-0.35 \leq  \frac{f_{WW}+f_{BB}-f_{BW}}{\Lambda^2} \leq 0.46
\;\;\;\; \mbox{ and } \;\;\;\;
2.8 \leq  \frac{f_{WW}+f_{BB}-f_{BW}}{\Lambda^2} \leq 3.6\; .
\label{lhclim}
\end{equation}
for $m_H\leq 150$ GeV.

The anomalous Higgs interaction $f_{BW}$ can also be constrained by
their effect on the triple gauge--boson vertices, but this is not
the case for $f_{WW}$ nor $f_{BB}$. 

In the following we will present our limits in terms of the relevant combination
$f = f_{WW}+f_{BB}-f_{BW} $ which is the only combination of anomalous couplings
directly measured in the processes
we study.

\section{Simulations}

In order to perform the Monte Carlo analysis, we have employed the
package MadGraph \cite{Madgraph} coupled to HELAS \cite{helas}.
Special subroutines were constructed for the anomalous contribution which
enable us to take into account all interference effects between
the QED and the anomalous amplitudes. The phase space integration
was performed by VEGAS \cite{vegas}.

The photon distribution in the nucleus can be described using the
equivalent-photon or Weizs\"{a}cker-Williams approximation in the impact
parameter space. Denoting the photon distribution function in a nucleus by
$F(x)$, which represents the number of
photons carrying a fraction between $x$ and $x+dx$ of the
total momentum of a nucleus of charge $Ze$, we can define
the two-photon luminosity through
\begin{equation}
\frac{dL}{d\tau} = \int ^1 _\tau \frac{dx}{x} F(x) F(\tau/x),
\end{equation}

\noindent
where $\tau = {\hat s}/s$, $\hat s$ is the square of the center
of mass (c.m.s.) system energy of the two photons and $s$ of the
ion-ion system. The total cross section
$ AA \rightarrow AA \gamma \gamma \rightarrow AA X$, where
$X$ are the particles produced by the $\gamma \gamma$ process, is
\begin{equation}
\sigma (s) = \int d\tau \frac{dL}{d\tau} \hat \sigma(\hat s),
\label{sigfoton}
\end{equation}
where $ \hat \sigma(\hat s)$ is the cross-section of the subprocess
$\gamma \gamma \rightarrow X$.

We choose to use the conservative and more realistic photon
distribution of Cahn and Jackson~\cite{cahn}, including a
prescription proposed by Baur~\cite{baur2} for realistic
peripheral collisions, where we must enforce that the minimum
impact parameter ($b_{min}$) should be larger than $R_1 + R_2$,
where $R_i$ is the nuclear radius of the ion $i$. A useful fit
for the two-photon luminosity is:
\begin{equation}
\frac{dL}{d\tau}=\left(\frac{Z^2 \alpha}{\pi}\right)^2 \frac{16}{3\tau}
\xi (z),
\label{e3}
\end{equation}
where $z=2MR\sqrt{\tau}$, $M$ is the nucleus mass, $R$ its radius and
$\xi(z)$ is given by
\begin{equation}
\xi(z)=\sum_{i=1}^{3} A_{i} e^{-b_{i}z},
\label{e4}
\end{equation}
which is a fit resulting from the numerical integration of the photon
distribution, accurate to $2\% $ or better for $0.05<z<5.0$, and where
$A_{1}=1.909$, $A_{2}=12.35$, $A_{3}=46.28$, $b_{1}=2.566$,
$b_{2}=4.948$, and $b_{3}=15.21$. For $z<0.05$ we use the expression (see
Ref.~\cite{cahn})
\begin{equation}
\frac{dL}{d\tau}=\left(\frac{Z^2 \alpha}{\pi}\right)^2
\frac{16}{3\tau}\left[\ln{\left(\frac{1.234}{z}\right)}\right]^3 .
\label{e5}
\end{equation}

We consider Ca-Ca collisions since they are the most promissing ones
to put limits on the anomalous couplings because of the larger
luminosity of the Ca beams. The energy for $^{40}_{20}$Ca considered 
was 140 TeV/beam with a luminosity of  $5 \times 10^{30}$ 
cm$^{-2}$ s$^{-1}$ at LHC \cite{calcio}. Since the collider will run 
in the heavy ion mode only a few months per year we will consider 
two possibilities for an integrated luminosity per year, one 
optmistic of 50 pb$^{-1}$ year$^{-1}$ and another more realistic of 
10 pb$^{-1}$ year$^{-1}$. 

\section{Results}

In our analyses we computed the SM and anomalous cross sections for
the Higgs production via photon-photon fusion in
peripheral heavy ion collisions at LHC using similar cuts and
efficiencies as the ones ATLAS Collaboration \cite{atlas} applied in
their studies of Higgs boson searches.

We begin our analyses imposing the following acceptance cuts
\begin{eqnarray}
p_{T}^{\gamma (b)} >  25 \; \text{GeV} \;\;\; ,
\;\;\;\;\;\;
|\eta_{\gamma (b)}| <   2.5      \;\;\; ,
\;\;\;\;\;\;
\Delta R_{\gamma \gamma (b\bar{b)}} >   0.7 \;\; ,
\label{cut1}
\end{eqnarray}
 and taking into account an  efficiency for reconstruction and
identification of one photon of 84\% and a b-tagging of 60\%
\cite{atlas}.

In order to improve our limits on the anomalous couplings, we
have studied several kinematical distributions of the final state
particles. The most promissing one is the invariant mass of the final
particles, since the anomalous interactions occur mainly for the Higgs
boson produced on-shell.

For instance, the number of SM cross section for the process $\gamma \gamma
\to b\bar{b}$ with  $m_H=115$ GeV falls from $\sim$25.4 pb  to
$\sim$4.06 pb when the cut $|m_{b \bar{b}} - m_H| < 15$ GeV is applied.
The pure anomalous cross section for $\gamma \gamma \to H \to b\bar{b}$
with $f=10$ TeV$^{-2}$ falls from 16.2 pb to 15.8 pb, being almost
unaffected by the invariant mass cut. The significance of a anomalous 
signal, given by ${\cal S} = N_{signal}/\sqrt{N_{SM}}$, is enhanced by
a factor of 2.4 when the invariant mass cut is used.

Therefore, for the photon-photon initial state, we collected the final
state $\gamma \gamma$ and/or $b \bar{b}$ events whose invariant masses
fall in bins of size of 15 GeV around the Higgs mass
\begin{eqnarray}
m_H - 15\text{ GeV}  < m_{\gamma \gamma (b \bar{b})}  < m_H + 15\text{ GeV}
\;\;
\label{cut2}
\end{eqnarray}
in order to evaluate our results.

Considering the set of cuts (\ref{cut1}) and (\ref{cut2}), the 
luminosity and efficiencies discussed above, and a Higgs mass in
the range (115--180) GeV, the number of Standard Model events for the 
process $\gamma \gamma \to \gamma \gamma$ is  smaller than one 
since the highest Standard Model cross section is smaller than 3 fb. 
Since we expect nearly zero events for this process, a 95\% CL limit 
for the anomalous couplings is obtained when its contribution
generates 3 events.

In Table \ref{cross_sec} we present the Standard Model cross section
for the process $\gamma \gamma \to b\bar{b}$ considering a Higgs mass 
in the range (115--180) GeV. For example, for a Higgs mass of 115 GeV,
the number of Standard Model events is $\sim$ 73(15) in one year when we
consider a luminosity of 50(10) pb$^{-1}$year$^{-1}$, including b-tagging
efficiency. In this case,
a 95\% CL signal is obtained when the number of SM events ($N_{SM}$)
is changed by a value of $2 \times \sqrt{N_{SM}}$ if $N_{SM}$ is greater 
than 10 units, otherwise we apply Poisson statistics for few
background events. 

In Tables \ref{fall_gamma} and \ref{fall_b} we present the limits for
$f$ considering the same range of Higgs masses. The limits are more
stringent  in the $\gamma \gamma \to b \bar{b}$ case, where the
number of events is larger. The limits get worse for $m_H > 160$ GeV
because the total Higgs width increases due to the opening of $W^+ W^-$
decay channel. 

\begin{figure}
\protect
\centerline{(a)}
\centerline{\mbox{\epsfig{file=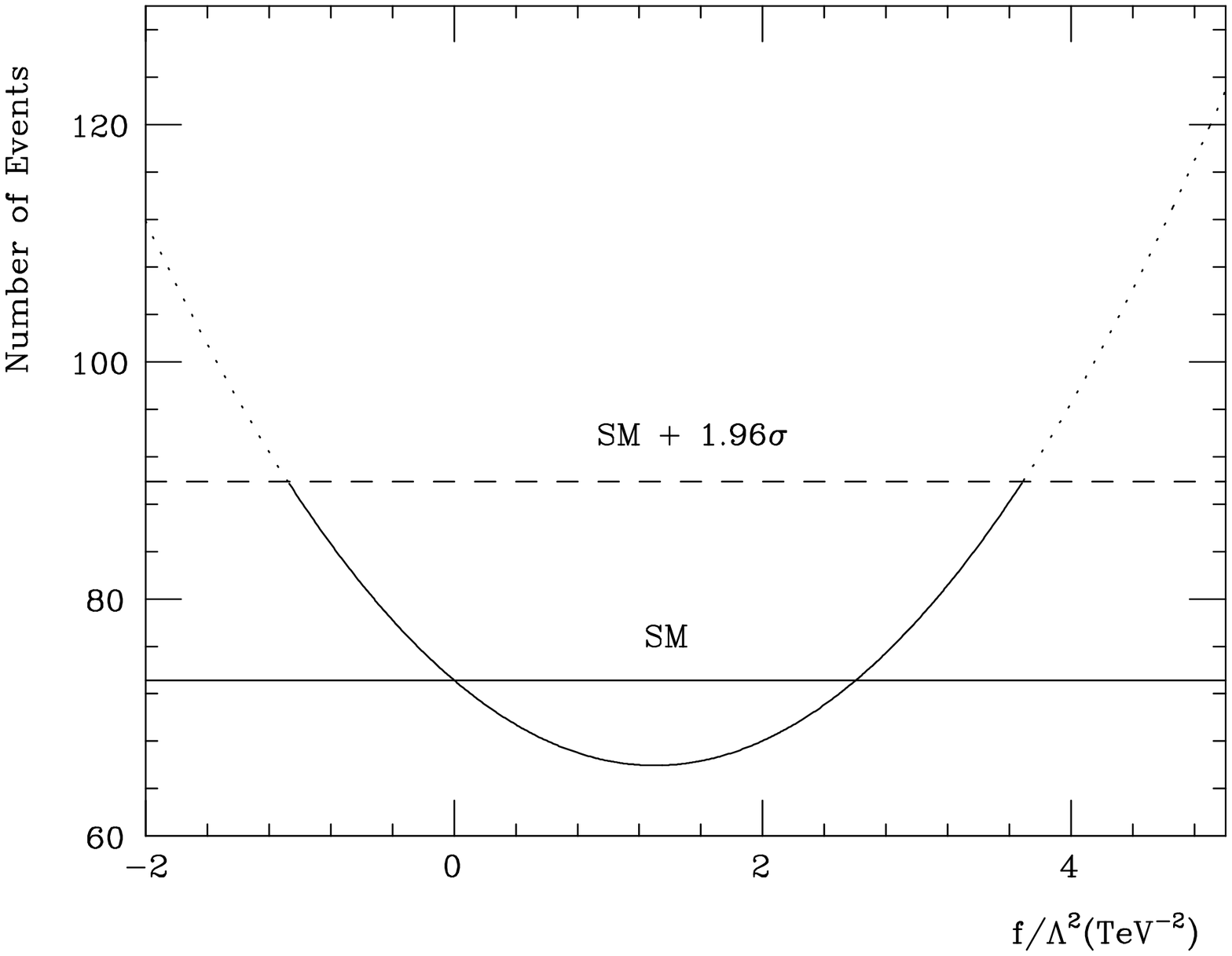,angle=90,width=0.6\textwidth}}}
\centerline{(b)}
\centerline{\mbox{\epsfig{file=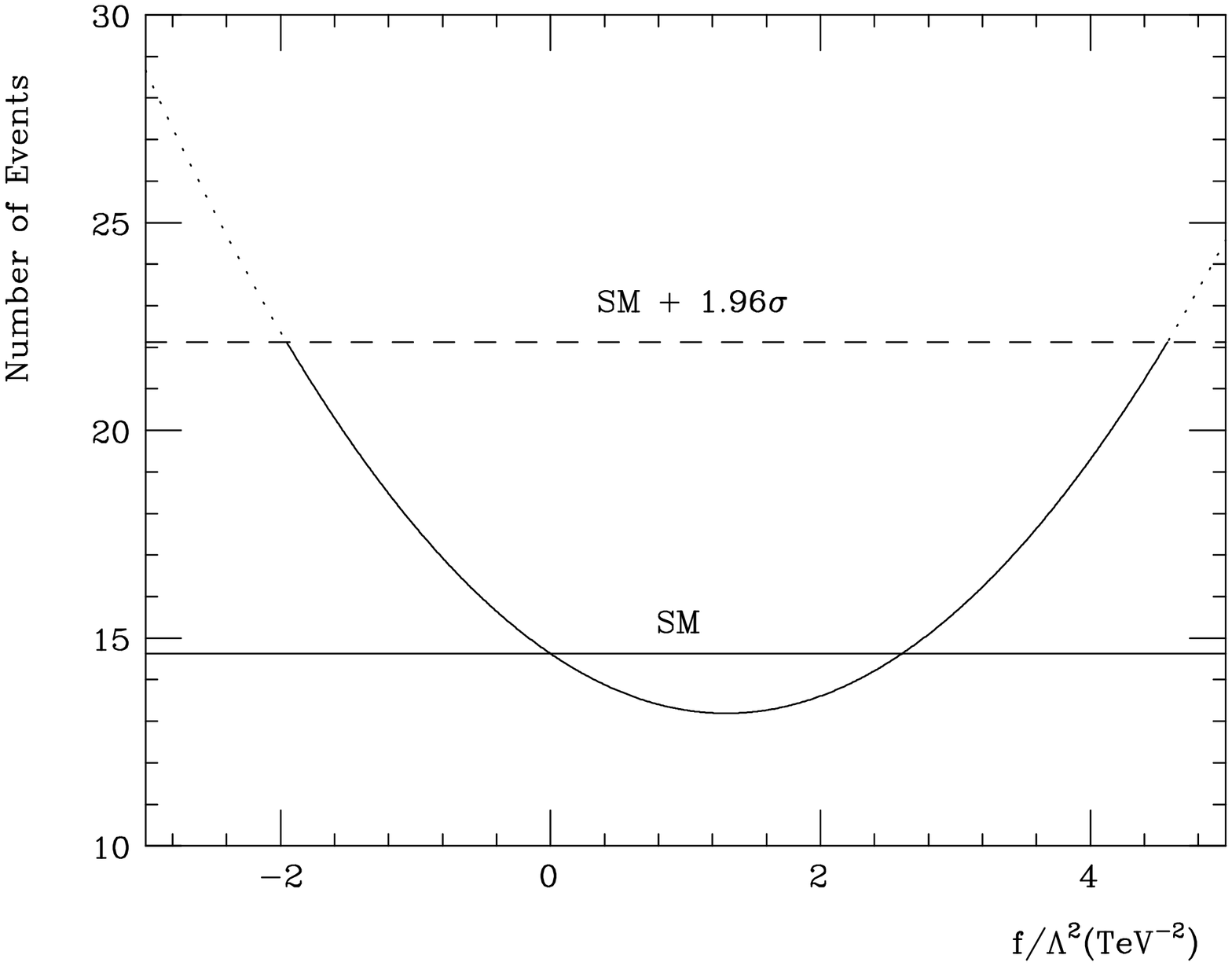,angle=90,width=0.6\textwidth}}}
\caption{Number of $b \bar{b}$ events in the LHC heavy ion mode 
with a luminosity of (a) 50 and (b) 10 pb$^{-1}$year$^{-1}$ as 
a function of the anomalous coupling $f$ for $m_H = 115$ GeV. 
The solid horizontal line is the number of events in the SM and the 
dashed horizontal line give the 95\% CL region. The solid part 
of the parabola represents the allowed region.}
\label{htb1}
\end{figure}

The pure anomalous contribution to the process $\gamma \gamma \to b
\bar{b}$ is quadratic in the anomalous coupling because there is only
one anomalous vertex in this case. In Figs. \ref{htb1}(a) and
\ref{htb1}(b) the number of $b \bar{b}$ events in the LHC heavy ion
mode as a function of the anomalous coupling $f$ is shown
together with the SM 95\% CL region for $m_H = 115$ GeV and a
luminosity of 50 and 10 pb$^{-1}$year$^{-1}$, respectively. 

\begin{figure}
\protect
\centerline{(a)}
\centerline{\mbox{\epsfig{file=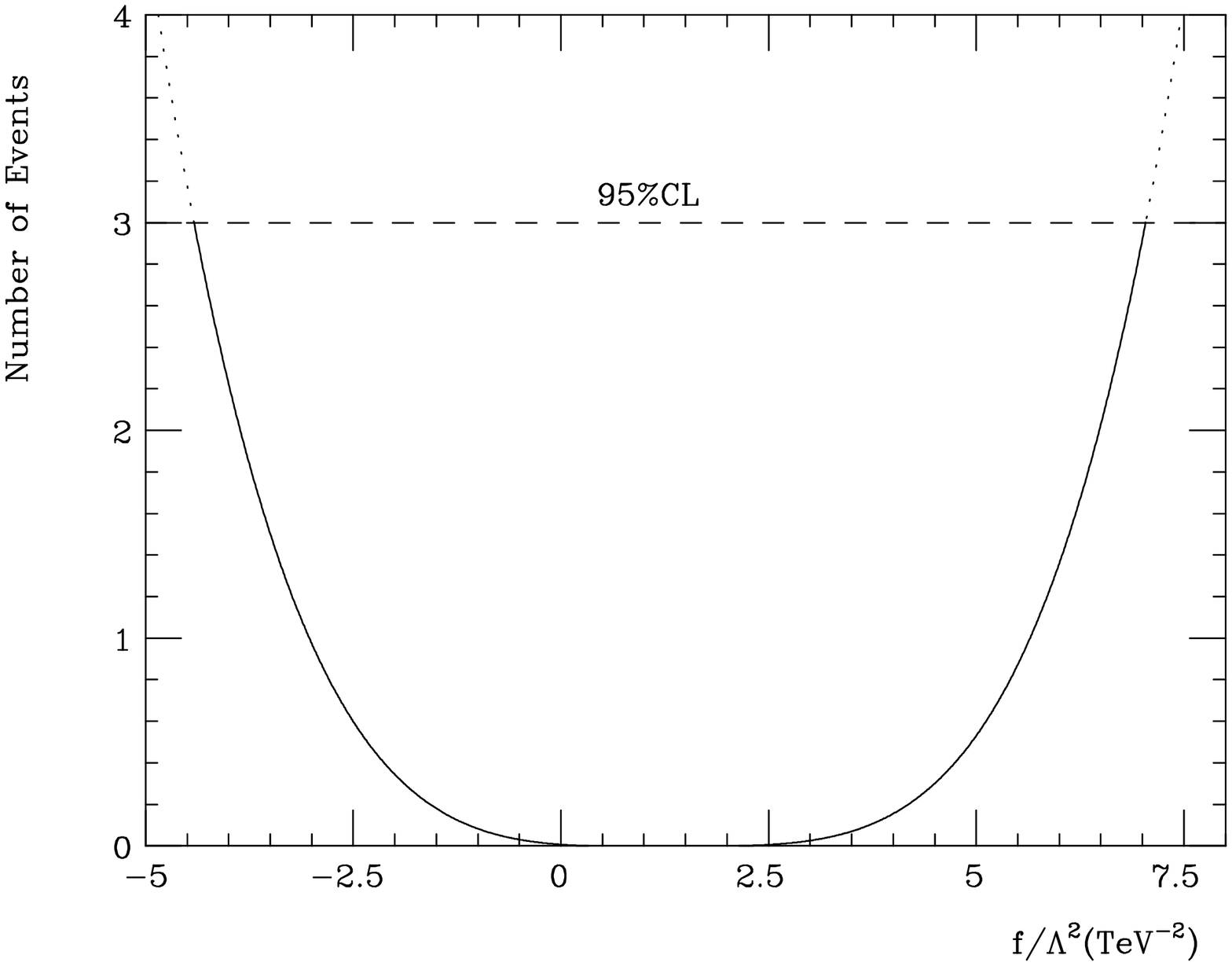,angle=90,width=0.6\textwidth}}}
\centerline{(b)}
\centerline{\mbox{\epsfig{file=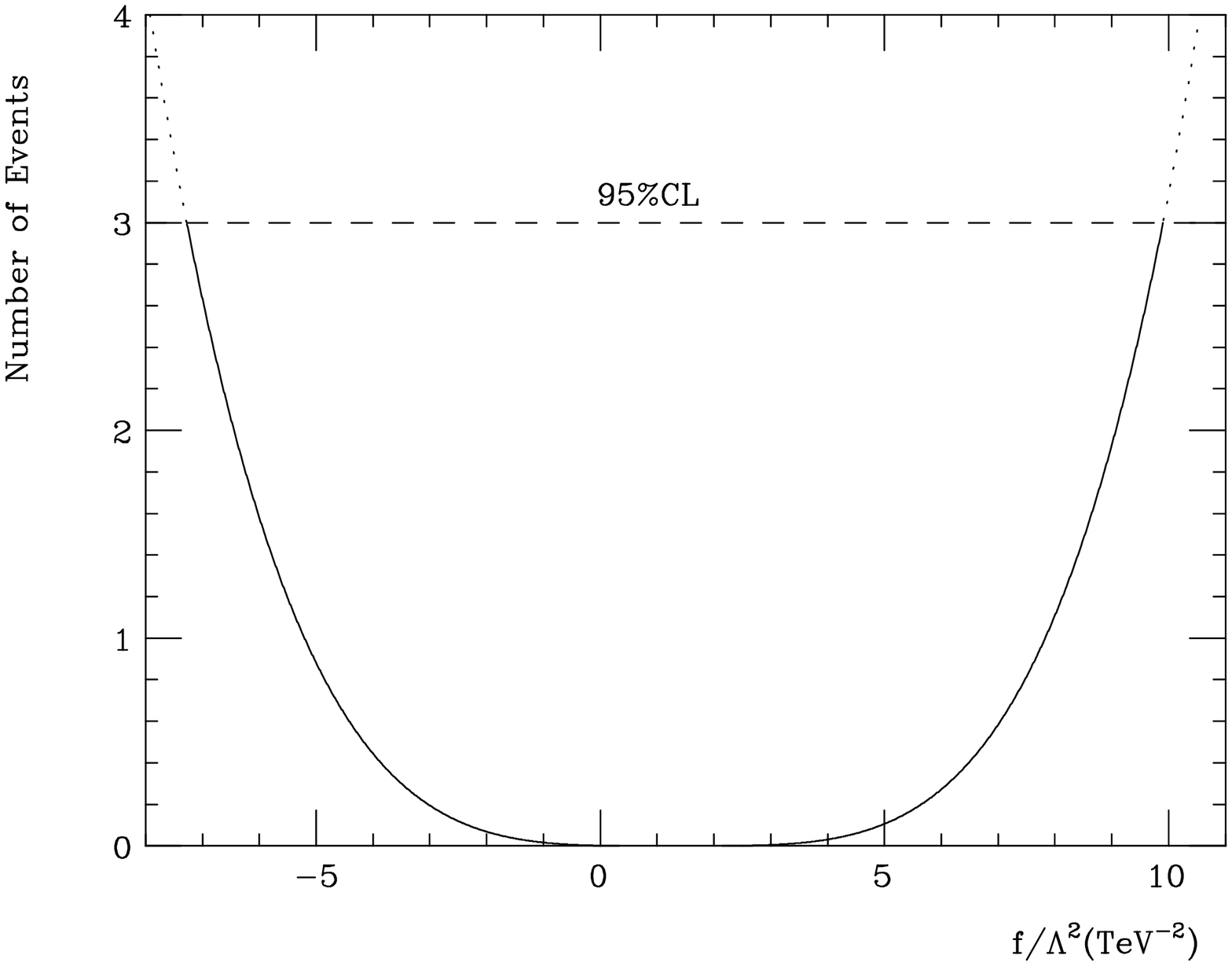,angle=90,width=0.6\textwidth}}}
\caption{Number of $\gamma \gamma$ events in the LHC heavy ion mode 
with a luminosity of (a) 50 and (b) 10 pb$^{-1}$year$^{-1}$ as 
a function of the anomalous coupling $f$ for $m_H = 115$ GeV. 
The solid horizontal line is the number of events in the SM and the 
dashed horizontal line give the 95\% CL region. The solid part 
of the curve represents the allowed region.}
\label{htb2}
\end{figure}

On the other hand, for the process $\gamma \gamma \to \gamma \gamma$,
the pure anomalous contribution is proportional to the fourth power
in the anomalous coupling because there are two anomalous vertices in 
this case, as shown in Figs. \ref{htb2}(a) and \ref{htb2}(b).

\section{Conclusions}
\label{sec:con}

In this work we have studied the sensitivity of the heavy ion mode of
the LHC to anomalous Higgs boson couplings to photons,
$H\gamma\gamma$, through the analysis of the processes
$\gamma \gamma \to b \bar{b},\gamma\gamma$
in peripheral heavy ion collisions.

Our best limits for the photon-photon initial state are (in TeV$^{-2}$),
\[
-1.1 (-2.0)\leq \frac{f}{\Lambda^2}\leq 3.7 (4.6)\;\;, \mbox{for} \;\;
\gamma \gamma \to b \bar{b}
\]
and
\[
-4.4 (-7.3)\leq \frac{f}{\Lambda^2}\leq 7.3 (9.9) \;\;, \mbox{for} \;\;
\gamma \gamma \to \gamma \gamma
\]
for an integrated luminosity 50 (10) pb$^{-1}$ year$^{-1}$, including $\gamma$
identification and b-tagging efficiencies.

These results are more stringent than the limits coming from the
proton--antiproton mode of the Tevatron 2.
We have also studied Higgs production via pomeron-pomeron fusion and found it
neglegible.

In conclusion, the limits for anomalous Higgs couplings that can
be obtained in peripheral heavy ion collisions at the LHC via
electromagnetic processes are a factor of five  tighter than
the limits that can be obtained in the upgraded Tevatron. However
the proton-proton mode of the LHC will be able to put stronger constraints due
to its  higher luminosity. 

\acknowledgments

This work was supported by Conselho Nacional de Desenvolvimento
Cient\'{\i}fico e Tecnol\'ogico (CNPq), by FINEP (PRONEX) and by 
Funda\c{c}\~ao
de Amparo \`a Pesquisa do Estado de S\~ao Paulo (FAPESP).


\widetext

\begin{table}
\begin{tabular}{||c||c||}
Higgs Mass(GeV) &
$\sigma(\gamma \gamma \to b \bar{b})$ (pb) \\
\hline
\hline
115 &  $4.06$ \\
\hline
120 &  $3.41$ \\
\hline
130 &  $2.44$ \\
\hline
140 &  $1.78$ \\
\hline
150 &  $1.34$ \\
\hline
160 &  $1.13$ \\
\hline
170 &  $0.53$ \\
\hline
180 &  $0.38$
\end{tabular}
\medskip
\caption{
Standard Model cross sections in pb for the process
$\gamma \gamma \to b \bar{b}$ for different Higgs boson masses
considerin the set of cuts (\protect{\ref{cut1}}) and
(\protect{\ref{cut2}}).}
\label{cross_sec}
\end{table}


\begin{table}
\begin{tabular}{||c||c||c||}
Higgs Mass(GeV) & ${\cal L} = 50$ pb$^{-1}$ & ${\cal L} = 10$ pb$^{-1}$ \\
\hline
\hline
115 & ($-$4.42,7.04) & ($-$7.28,9.90) \\
\hline
120 & ($-$4.41,7.09) & ($-$7.28,9.97) \\
\hline
130 & ($-$4.41,7.24) & ($-$7.31,10.2) \\
\hline
140 & ($-$4.39,7.44) & ($-$7.36,10.4) \\
\hline
150 & ($-$4.34,7.73) & ($-$7.36,10.8) \\
\hline
160 & ($-$4.04,8.40) & ($-$7.14,11.5) \\
\hline
170 & ($-$13.0,16.9) & ($-$20.4,24.3) \\
\hline
180 & ($-$15.0,18.7) & ($-$23.3,27.0) 
\end{tabular}
\medskip
\caption{95 \% CL allowed regions for $f$ in TeV$^{-2}$ 
for different Higgs boson masses in the process 
$\gamma \gamma \to H \to \gamma \gamma$.}
\label{fall_gamma}
\end{table}


\begin{table}
\begin{tabular}{||c||c||c||}
Higgs Mass(GeV) & ${\cal L} = 50$ pb$^{-1}$ & ${\cal L} = 10$ pb$^{-1}$ \\
\hline
\hline
115 &  ($-$1.08,3.69) & ($-$1.96,4.57) \\
\hline
120 &  ($-$1.08,3.74) & ($-$1.95,4.63) \\
\hline
130 &  ($-$1.07,3.87) & ($-$2.17,4.99) \\
\hline
140 &  ($-$1.06,4.06) & ($-$2.20,5.23) \\
\hline
150 &  ($-$1.05,4.38) & ($-$2.34,5.73) \\
\hline
160 &  ($-$1.01,5.22) & ($-$2.35,6.66) \\
\hline
170 &  ($-$10.9,29.2) & ($-$19.2,34.8) \\
\hline
180 &  ($-$14.7,26.8) & ($-$51.5,32.5)
\end{tabular}
\medskip
\caption{95 \% CL allowed regions for $f$ in TeV$^{-2}$ 
for different Higgs boson masses in the process 
$\gamma \gamma \to H \to b \bar{b}$.}
\label{fall_b}
\end{table}

\newpage

\newpage

\begin{references}

\bibitem{higgs} See
{\tt http://delphiwww.cern.ch/$\sim$offline/physics\_links/lepc.html}.

\bibitem{baur}
G.\ Baur, J.\ Phys.\ {\bf G24} (1998) 1657 .

\bibitem{baur2}
G.\ Baur, K.\ Hencken and D.\ Trautmann, hep-ph/9810418;
C.\ A.\ Bertulani and G.\ Baur, Phys.\ Reports \ {\bf 163} (1988) 299 ;
G.\ Baur, in {\it Proceedings of the CBPF International Workshop on
Relativistic Aspects of Nuclear Physics}, Rio de Janeiro, 1989, edited
by T.~Kodama {\it et al.} (World Scientific, Singapore,  1990), p. 127;
G.\ Baur and C.\ A.\ Bertulani, Nucl.\ Phys.\ {\bf A505} (1989) 835.

\bibitem{papa}
E.\ Papageorgiu, Phys.\ Rev.\ {\bf D40} (1989) 92; Nucl.\  Phys.\
{\bf A498} (1989) 593c;
M.\ Grabiak {\it et al.}, J.\  Phys.\ {\bf G15} (1989) L25;
M.\ Drees, J.\ Ellis and D.\ Zeppenfeld, Phys.\ Lett.\ {\bf B223}
(1989) 454;
M.\ Greiner, M.\ Vidovic, J.\ Rau and G.\ Soff, J.\ Phys.\ {\bf G17}
(1991) L45;
B.\ M\"uller and A.\ J.\ Schramm, Phys.\ Rev.\ {\bf D42} (1990) 3699;
J.\ S.\ Wu, C.\ Bottcher, M.\ R.\ Strayer
and A.\ K.\ Kerman, Ann.\ Phys.\ {\bf 210} (1991) 402.

\bibitem{cahn} R.\ N.\ Cahn and J.\ D.\ Jackson,
Phys.\ Rev.\ {\bf D42} (1990) 3690.

\bibitem{muller} B.\ M\"uller and A.\ J.\ Schramm, Nucl.\ Phys.\
{\bf A523} (1991) 677.

\bibitem{effective} S.\ Weinberg, Physica {\bf 96A} (1979) 327;
J.\ F.\ Donoghue, E.\ Golowich and B.\ R.\ Holstein, {\it
Dynamics of the Standard Model} (Cambridge University Press,
1992).

\bibitem{linear} W.\ Buchm\"uller and D.\ Wyler, Nucl.\ Phys.\ {\bf
B268} (1986) 621; C.\ J.\ C.\ Burgess and H.\ J.\ Schnitzer, Nucl.\
Phys.\ {\bf B228} (1983) 454; C.\ N.\ Leung, S.\ T.\ Love, and S.\
Rao, Z.\ Phys.\ {\bf 31} (1986) 433.

\bibitem{dim6:zep} K.\ Hagiwara, S.\ Ishihara, R.\ Szalapski and
D.\ Zeppenfeld, Phys.\ Rev.\ D {\bf 48} (1993) 2182.

\bibitem{Hagiwara2} K.\ Hagiwara, R.\ Szalapski, and D.\
Zeppenfeld, Phys.\ Lett.\ {\bf B318} (1993) 155.

\bibitem{rujula}  A.\ De R\'ujula, M.\ B.\ Gavela, P.\ Hern\'andez,
and E.\ Mass\'o, Nucl.\ Phys.\ {\bf B384} (1992) 3.

\bibitem{hms} K.\ Hagiwara, S.\ Matsumoto, and R.\ Szalapski, Phys.\
Lett.\ {\bf B357} (1995) 411; S.\ Alam, S.\ Dawson, and R.\
Szalapski, Phys.\ Rev.\ {\bf D57} (1998) 1577.

\bibitem{ours:lep2aaa}  O.\ J.\ P.\ \'Eboli, M.\ C.\
Gonzalez--Garcia,  S.\ M.\ Lietti, and S.\ F.\ Novaes, Phys.\ Lett.\
{\bf B434} (1998) 340.

\bibitem{ours:tevatron} F.\ de Campos, M.\ C.\ Gonzalez--Garcia, and
S.\ F.\ Novaes, Phys.\ Rev.\ Lett.\ {\bf 79} (1997) 5210; M.\ C.\
Gonzalez--Garcia, S.\ M.\ Lietti, and S.\ F.\ Novaes, Phys.\ Rev.\
{\bf D57} (1998) 7045; F.\ de Campos, M.\ C.\ Gonzalez--Garcia, S.\
M.\ Lietti, S.\ F.\ Novaes, and R.\ Rosenfeld, Phys.\  Lett.\ {\bf
B435} (1998) 407.

\bibitem{ours:comb} M.\ C.\ Gonzalez--Garcia, S.\ M.\ Lietti, and S.\
F.\ Novaes, Phys.\ Rev.\ {\bf D59} (1999) 075008.

\bibitem{ours:lhc} O.\ J.\ Eboli, M.\ C.\ Gonzalez-Garcia,
S.\ M.\ Lietti and S.\ F.\ Novaes, Phys.\ Lett.\  {\bf B478} (2000) 199.

\bibitem{Madgraph} T.\ Stelzer and W.\ F.\ Long, Comput.\
Phys.\ Commun.\ {\bf 81} (1994) 357.

\bibitem{helas} H.\ Murayama, I.\ Watanabe and K.\ Hagiwara,
KEK Report 91-11 (unpublished).

\bibitem{vegas} G.\ P.\ Lepage,  J. \ Comp. \ Phys. {\bf 27}
(1978) 192, and ``Vegas: An Adaptive Multidimensional Integration
Program", CLNS-80/447, 1980 (unpublished).

\bibitem{calcio} E.\ Papageorgiu, hep-ph/9507221.

\bibitem{atlas} Atlas Detector and Physics Performance Technical 
Design Report, \\
{\tt http://atlasinfo.cern.ch/Atlas/GROUPS/PHYSICS/TDR/access.html}.

\end{references}
\end{document}